\documentclass{amsart}
\RequirePackage{fix-cm}

\usepackage{wasysym}
\usepackage{amssymb}

\usepackage{natbib}
\usepackage[margin=1.2in]{geometry}

\newcommand{\norm}[1]{\left\lVert#1\right\rVert}
\newcommand{\ie}{\textit{i.e.,\ }}
\newcommand{\eg}{\textit{e.g.,\ }}
\newcommand{\dd}{\textrm{d}}

\newcommand{\DLD}{\norm{\Delta \textrm{LD}}}
\newcommand{\DD}{\norm{\Delta D}}

\usepackage{enumitem}   

\usepackage{rotating}

\usepackage[colorlinks,citecolor=blue]{hyperref}
\begin{document}

\title[Lagrangian descriptors for chaos detection]{
	Lagrangian descriptors and their applications to deterministic chaos
}

\author[J.\,Daquin]{
J\'er\^ome Daquin         
}
\address{
	Department of Mathematics and naXys, Namur Institute for Complex Systems,
	University of Namur, Rue Grafé 2, B5000 Namur, Belgium}
\email{jerome.daquin@unamur.be}

\date{\today}

\date{\today}

\begin{abstract}
	We present our recent contributions to the theory of Lagrangian descriptors for discriminating ordered and deterministic chaotic trajectories. The class of Lagrangian Descriptors we are dealing with is based on the Euclidean length of the orbit over a finite time window. 
	The framework  is free of tangent vector dynamics and is valid for both discrete and continuous dynamical systems. 
	We review its last advancements  and touch on how it illuminated recently Dvorak's quantities based on maximal extent of trajectories' observables, as traditionally computed in planetary dynamics.
\end{abstract}
\maketitle

\tableofcontents
\section{Introduction}
This present short presentation aims at summarising our recent  contributions to the theory of Lagrangian Descriptors (LDs) for deterministic chaos detection. It is predominantly based on \cite{jDa22-physD} and \cite{jDa23} to which we refer for a comprehensive literature review.  
LDs are scalar quantities rooted in fluid mixing and coined as such by \cite{cMe10}. There were initially introduced by \cite{jaMa09} for vector fields.
Consider the autonomous regular vector field  
\begin{align}\label{eq:VectorField}
	\dot{x}=v(x), \, x \in \mathbb{R}^{n},
\end{align}
and the orbit associated to Eq.\,(\ref{eq:VectorField}) starting at $x_{0} \in \mathbb{R}^{n}$ at $t=0$. 
Its LD at time $T > 0$ is defined as 
\begin{align}
	\textrm{LD}(x_{0};T) = 
	\int_{0}^{T} \norm{v\big(x(s)\big)}_{2} \, \dd s,
\end{align}
and corresponds to the Euclidean arc-length of the trajectory over the time-window $[0,T]$. 
LDs have been generalised by considering the integration of others bounded and intrinsic observable associated to an orbit \citep{aMa13}. 
We stick here to the LD based on the arc-length method. Extending LDs to discrete systems is straightforward.   
Let us denote by $\{z_{0},z_{1},\dots,z_{N}\}$, 
$z_{j} \in \mathbb{R}^{m}\, j=0,\cdots,N$, a finite orbit associated to a discrete mapping $f$, $z_{j}=f^{j}(z_{0}), j=1,\cdots,N$.
The LD associated to such orbit is given by 
\begin{align}\label{eq:LD}
	\textrm{LD}(z_{0};N) =
	\sum_{j=0}^{N-1}
	\sqrt{\sum_{l=1}^{m}
		(z_{j+1}^{(l)}-z_{j}^{(l)})^{2}},
\end{align}
where $z_{j}^{(l)}$ denotes the $l$-th component of the state $z$ at time $j$.  \\

The rest of this contribution is organised as follows. 
In Sect.\,\ref{sec:LDLogisticModel},  we highlight the phenomenology of the LDs for chaos detection by presenting new applications  to the paradigmatic one-dimensional logistic map. 
Sect.\,\ref{sec:LDChaos} presents the finite-time chaos indicator we have derived from LDs computations and discuss the models on which it has been successfully employed so far.  
Sect.\,\ref{sec:LDDvorak} connects the LD framework with maximal excursions traditionally computed in celestial mechanics in the context of stability maps. 
Sect.\,\ref{sec:ccl} closes the paper with conclusive remarks. 
\section{Lagrangian Descriptors applied to the logistic model}\label{sec:LDLogisticModel}
\begin{figure}
	\begin{center}
		\includegraphics[width=0.8\linewidth]{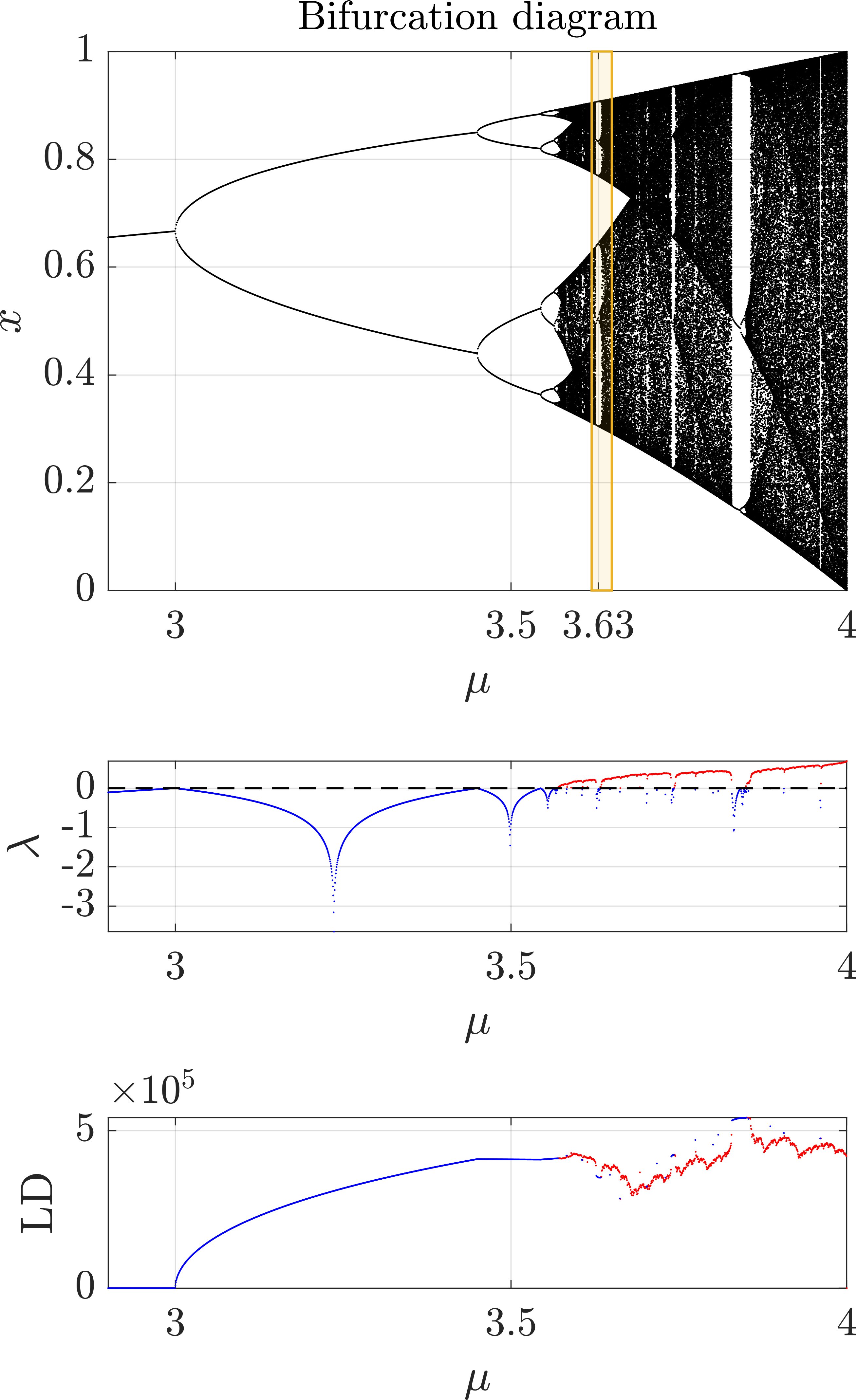}
		\caption{Bifurcation diagram (top panel), Lyapunov exponents (middle panel) and LD lengths (bottom panel) of the logistic equation when $\mu$ is varied. There is a strong connection between the regularity of the orbits (sign of Lyapunov exponent) and the regularity of the LD map. See text for further details.}
		\label{fig:fig1}
	\end{center}
\end{figure}

\begin{figure}
	\begin{center}
		\includegraphics[width=0.9\linewidth]{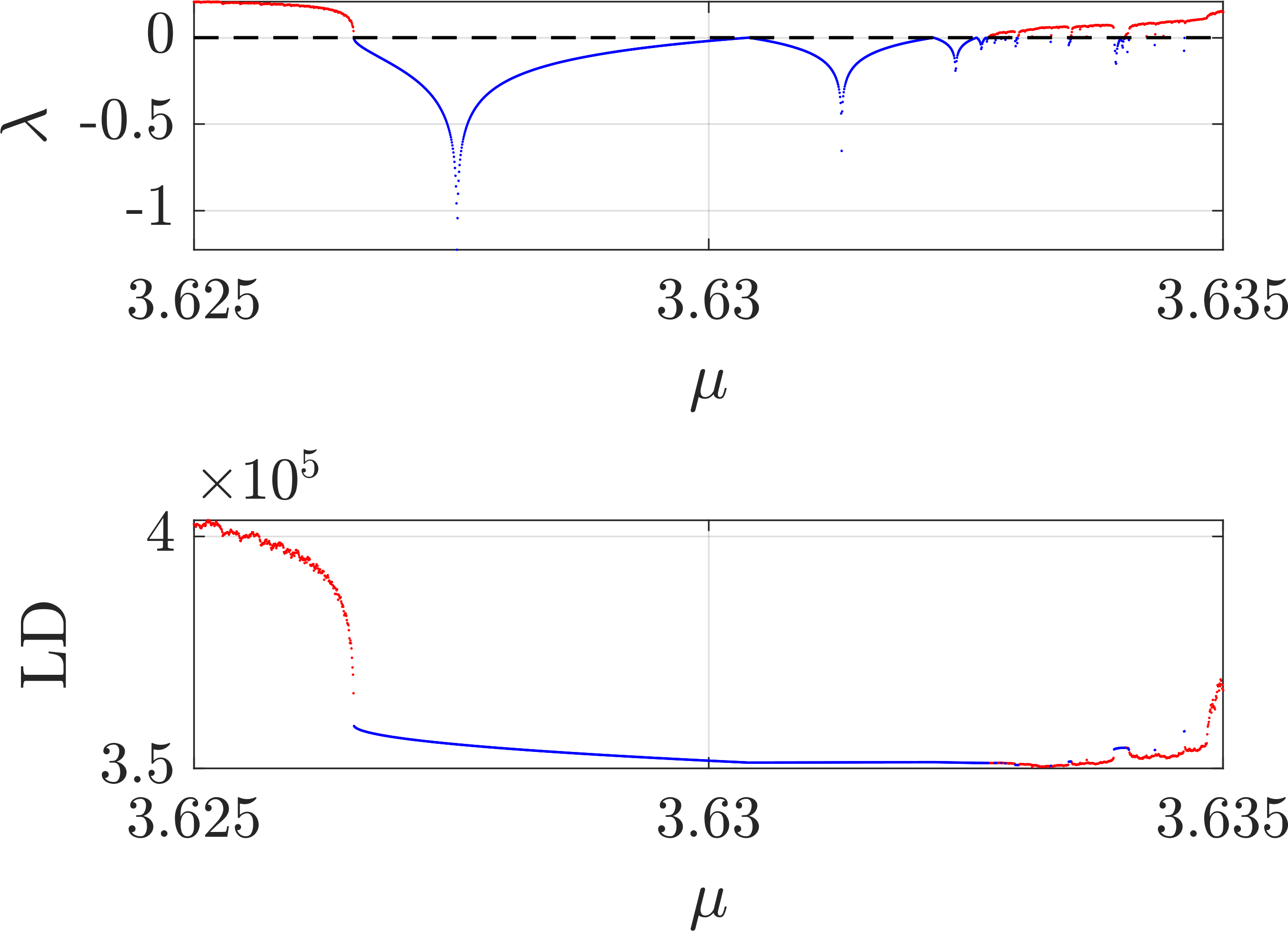}
		\caption{Computation of Lyapunov exponents $\lambda$ and LD lengths on a small range of the control parameter $\mu$. 
			The regularity of the length application provides a very sensitive and reliable indicator of the ordered and chaotic nature of the motion.}
		\label{fig:fig2}
	\end{center}
\end{figure}

\begin{figure}
	\begin{center}
		\includegraphics[width=0.6\linewidth]{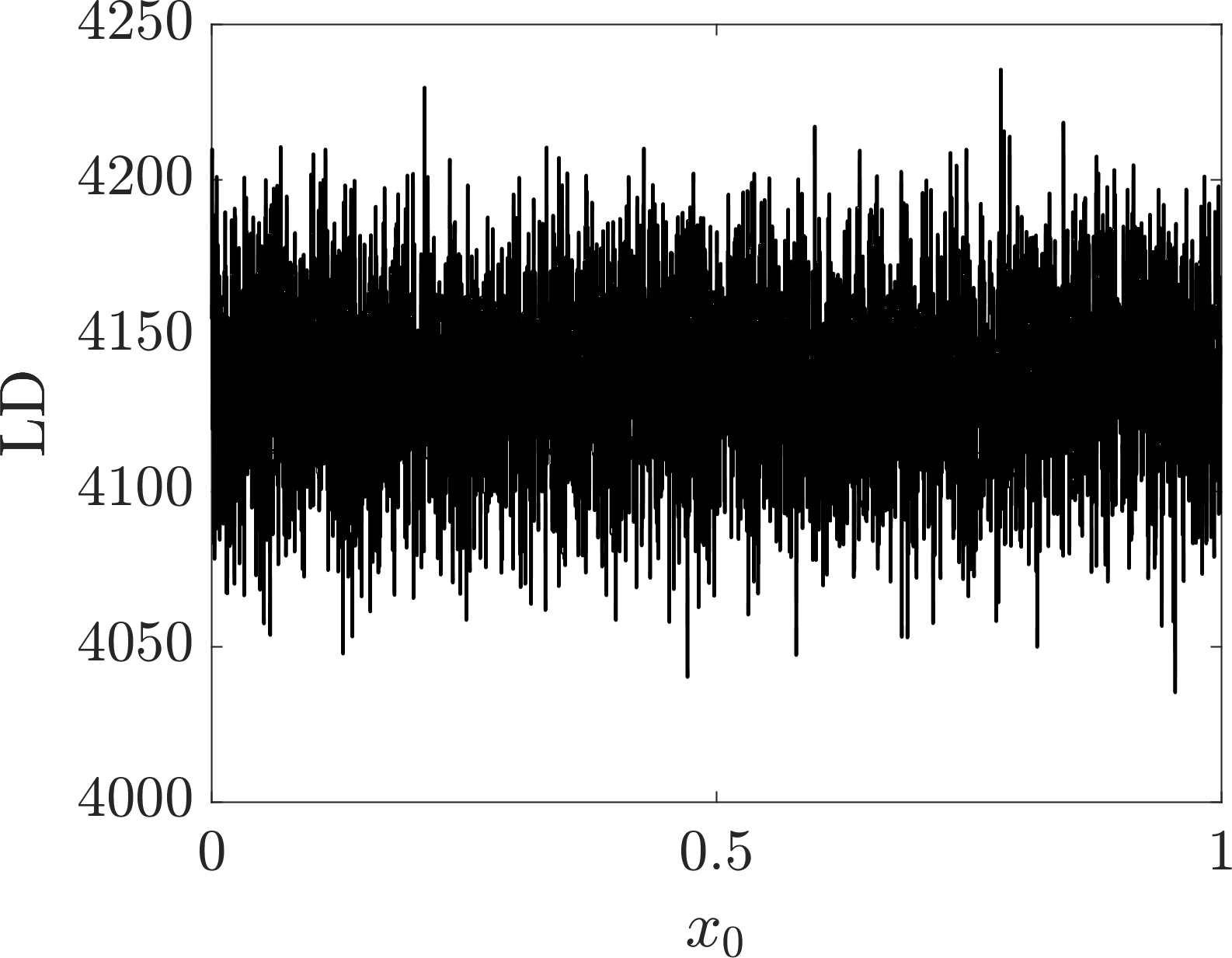}
		\caption{Landscape of LDs as a function of $x_{0}$ for the logistic map with $\mu=4$. For this parameter, the map is known to be chaotic. This is reflected in the absence of regularity of the LD metric.  
		}
		\label{fig:fig3}
	\end{center}
\end{figure}

We consider the one-dimensional quadratic logistic map (see \eg \cite{rMa76}) as a testing ground to illustrate the LDs concepts. 
The mapping is defined by
\begin{align}
	x_{n+1}=\mathcal{L}_{\mu}(x_{n})=\mu x_{n}(1-x_{n}).
\end{align} 
The state space of the dynamics is $M=[0,1]$ for $0< \mu\le 4$ and $n \in  \mathbb{N}$. 
Dealing with the orbit $\{x_{0},x_{1},\dots,x_{N}\}$, Eq.\,(\ref{eq:lengthDiscret}) becomes 
\begin{align}\label{eq:lengthDiscret}
	\textrm{LD}(x_{0};N)=
	\sum_{i=0}^{N-1}
	\vert x_{i+1}-x_{i} \vert. 
\end{align}
The top panel of Fig.\,\ref{fig:fig1} shows the bifurcation diagram of the logisitc map for $2.9 \le \mu \le 4$ with fixed $x_{0}=0.15$. It highlights the presence of periodic orbits, the chain of bifurcation occurring in the dynamics as $\mu$ is varied, 
periodic windows immersed within 
aperiodic  range of motions. 
Such a window is exemplified around 
$\mu=3.63$ and will be further discussed. 
The middle panel of Fig.\,\ref{fig:fig1} shows the value of
the  finite time Lyapunov exponent (FTLE) computed at time $N=10^{6}$ as a function of $\mu$. The FTLE is 
defined as
\begin{align}
	\lambda(x_{0};N)
	=
	\frac{1}{N} \log_{10}\big(
	\prod_{i=0}^{N}
	\vert D \mathcal{L}_{\mu}(x_{i})\vert 
	\big)
	=
	\frac{1}{N}\sum_{i=0}^{N}
	\log_{10}(\vert \mu (1-2x_{i})\vert).
\end{align}
We have colored the final value of $\lambda$ according to its sign. 
Negative FTLEs appear in blue and correspond to regular motions. 
The values $\lambda=0$ indicate bifurcations. 
The positive FTLEs appear in red and correspond to orbits with sensitive dependence upon the initial conditions, \ie chaotic motions. 
Interestingly enough, there is a sharp link between the values of the FTLEs and properties of the LD map. 
In fact, the bottom panel of Fig.\,\ref{fig:fig1} shows the LD landscape (computed also at the final time $N=10^{6}$), colored according to the final value of $\lambda$. We observe that negative $\lambda$'s correspond to smooth parts of the LD curve, whilst positive $\lambda$'s
correspond to domains where the LD metric is irregular.  
Fig.\,\ref{fig:fig2} repeats the computations of FTLEs and LDs at a much thinner scale of the control parameter $\mu$ (the shaded area around $\mu=3.63$ highlighted in Fig.\,\ref{fig:fig1}) and further confirm our former observations. 
This demonstrates the sensitivity and robustness of the regularity of the LD metric as diagnostic for chaos detection. 
For $\mu=4$, $\mathcal{L}_{4}$ is chaotic for every $x_{0}$ \citep{rMa76,jBa03}. 
Fig.\,\ref{fig:fig3} shows the LD landscape for varying $x_{0}$ in $(0,1)$ with fixed $\mu=4$ at $N=10^{4}$. The obtained landscape is nowhere smooth, in accordance with the former numerical results.   

\section{The new non-variational geometrical chaos indicators}\label{sec:LDChaos}
As just illustrated on the logistic model, the regularity of the LD metric keeps trace of the possible chaotic nature of the orbit. 
This property has also been observed on a series of integrable and non-integrable mechanical models supporting interacting resonances \citep{jDa22-physD}. 
For integrable 1-degree-of-freedom (DoF) systems, 
\cite{rPO22} showed that the rate of divergence of the derivative of the length metric (using a time-free parametrisation of LDs) scales as a power law when crossing transversally separatrices. The scaling obeys $\mathcal{O}(1/\sqrt{E})$, where $E$ is the energy labelling the level curve.  
Those observations led us to assume that the LD map is $\mathcal{C}^{\infty}$ on the set $\mathcal{R}$ of regular motions. 
Leveraging on this, \cite{jDa22-physD} introduced the 
$\DLD$ index measuring the regularity of the LD metric through second-derivatives estimates\footnote{
	An index based on the first derivatives might miss the geography of resonant webs, see \cite{jDa22-physD}, Fig.\,7.
}, similarly to the frequency analysis method \citep{jLa93}. 
For the dimension $n=1$ and a point $y \in \mathcal{R}$, $\DLD$ is defined as 
\begin{align}\label{eq:DLD1d}
	\DLD(y;T)=\vert \textrm{LD}^{''}(y;T)\vert.
\end{align}
Numerically, the second derivative in Eq.\,(\ref{eq:DLD1d}) is estimated by finite differences of the type 
\begin{align}
	\vert \textrm{LD}^{''}(y;T)\vert
	\simeq
	\frac{
		\vert 
		\textrm{LD}(y+h;T)+\textrm{LD}(y-h;T)-2
		\textrm{LD}(y;T)
		\vert}{h^{2}},
\end{align}
for a small enough discretisation step $h$. The $\DLD$ index is proposed as new chaos indicator and 
undergoes sharp increases on the complement set $\mathcal{R}^{C}$ of $\mathcal{R}$ when crossing transversally separatrices of integrable model or hyperbolic domains of non-integrable models \citep{jDa22-physD}.  
In the more general case ($n > 1$), $\DLD$ reads as
\begin{align}\label{eq:DLD}
	\DLD(y;T) = \sum_{i=1}^{n}
	\vert 
	\partial_{y^{(i)}y^{(i)}}^{2} \textrm{LD}(y;T)
	\vert. 
\end{align}
The $\DLD$ index has been applied and benchmarked on several low-dimensional models to derive stability maps computed on fine domains of initial conditions or parameter space, including: the $2$ dimensional standard map and higher $4$ dimensional coupled standard maps, non-autonomous pendulum-like models having resonant junctions and interactions supporting resonant webs, as ubiquitous in celestial mechanics. 
The $\DLD$ keeps trace of manifolds' oscillations when computed on a short timescale, as  observed on the
$2$-DoF H\'enon-Heiles model \citep{jDa22-physD}. 
Qualitative comparison with established variational methods (\eg the FLI, MEGNO, SALI) have shown excellent agreements in disentangling ordered and chaotic orbits. 
Recently, quantitative oriented analysis of the performance of LD like diagnostics have been investigated \citep{mHi22,sZi23}. 
On $4D$ coupled standard maps in a mixed phase space regime, it has been shown that the probability of agreement $P_{A}$ in the discrimination of the orbit against the SALI index is on the order of $90\%$. Thus, the $\DLD$ index is a cheap, easily implementable and reliable tool for detecting separatrices and chaotic motions.

\section{Diameters and Dvorak's amplitudes}\label{sec:LDDvorak}
The LDs just presented are based on the length of the orbit, see Eq.\,(\ref{eq:LD}) and (\ref{eq:lengthDiscret}). For bounded orbits, the amplitude (or diameter)
is another geometrical quantity that might be considered.
Consider Eq.\,(\ref{eq:VectorField}), a trajectory $x(t;x_{0})$ with initial condition $x_{0}$ 
and an observable $\phi$. The diameter on the finite time segment $[0,T]$ along the observable $\phi$
is defined as
\begin{align}\label{eq:D}
	D(x_{0};T)=
	\max_{0 \le \tau \le T} \phi\big(x(\tau;x_{0})\big) 
	-
	\min_{0 \le \tau \le T} \phi\big(x(\tau;x_{0})\big), 
\end{align}
with similar definitions for the discrete case. 
The diameter $D$ is called \textit{maximal shift} in \cite{rMu14}. The study of its cumulative distribution function demonstrated relevance to 
characterise fluid mixing properties, and is certainly an interesting direction of future research.   
\cite{jDa23} considered diameters of $n$-DoF Hamiltonian systems by looking more specifically at the diameter of the actions. It turns out that
LD and D maps share analogies, and of  practical interest is the loss of regularity when crossing separatrices or chaotic layers transversally.  
This led to introduce the analogue of Eq.(\ref{eq:DLD}) for the diameter of Eq.\,(\ref{eq:D}), denoted $\DD$.
Diameters like quantities have been computed for a while in planetary dynamics in the context of stability maps, by focusing on stretches of some keplerian elements $a$, $e$ or $i$ (respectively semi-major axis, eccentricity and inclination), usually denoted by $\delta a$, $\delta e$ or $\delta i$, see \eg \cite{rDv04,zSa07}.  
In the context of $2$ and $3$-body co-planar simulations triggered towards mean-motion resonances,  \cite{jDa23} shown in particular how the 
$\DD$ maps supplements the traditional diameter  analysis, allowing to reinflate and recover sharply separatrices, detect thin chaotic lines and the web of resonances, otherwise undetected with the classical diameter metric. As for $\DLD$, $\DD$ is a cheap non-variational geometrical index allowing sensitive chaos detection.

\section{Conclusive remarks}\label{sec:ccl}
This short manuscript has  presented and summarised our latest contributions to the theory of Lagrangian Descriptors for chaos detection. 
A finite-time non-variational chaos indicator can be easily derived from the the study of the regularity of length map. For this, we suggested  a second-derivatives based index.    
We have discussed connection of Lagrangian diagnostics with 
quantities employed to characterise fluid flow mixing and
diameter quantities encountered in celestial mechanics.  
The Lagrangian framework offers a convenient mold for detecting chaos without the need of deriving the variational equations, possibly leading to substantial implementation reduction. Our current efforts focus on the quantitative assessments of their performances.


\bibliographystyle{apalike} 
\bibliography{biblio}

\end{document}